\documentclass[usenatbib]{mn2e}
\usepackage{times}
\usepackage{txfonts}
\usepackage{epsfig}
\usepackage{graphicx}
\usepackage[margin=0.5in, top=1in, bottom=1in, a4paper, pdftex, dvips]{geometry}

\title[Variations in rotational frequency splittings]
{Misleading variations in estimated rotational frequency splittings of solar p modes: Consequences for helio- and asteroseismology}
\author[A.-M. Broomhall et al.]{Anne-Marie
Broomhall,$^1$\thanks{amb@bison.ph.bham.ac.uk} David
Salabert,$^{2, 3, 4}$ William J. Chaplin,$^1$ Rafael A. Garc{\'{i}}a,$^5$ \newauthor
Yvonne
Elsworth,$^1$ Rachel Howe$^1$ and Savita
Mathur$^6$\\ $^1$School of Physics and
Astronomy,
University of Birmingham, Edgbaston, Birmingham B15 2TT\\
$^2$Instituto de
Astrof{\'{i}}sica de Canarias, E-38200 La Laguna, Tenerife, Spain\\
$^3$Departamento do Astrof{\'{i}}sica, Universidad de La Laguna,
E-38206 La Laguna, Tenerife, Spain\\
$^4$Universit{\'e} de Nice Sophia-Antipolis, CNRS UMR 6202,
Observatoire de la C{\^{o}}te dAzur, BP 4229,
06304 Nice Cedex 4, France\\
$^5$Laboratoire AIM,
CEA/DSM-CNRS-Universit{\'e} Paris Diderot, IRFU/SAp, Centre de
Saclay, 91191 Gif-sur-Yvette, France\\
$^6$High Altitude Observatory,
NCAR, PO Box 3000, Boulder, CO 80307, USA}
\date{}

\begin{document}
\maketitle
\begin{abstract}The aim of this paper is to investigate whether there are any 11-yr or quasi-biennial solar cycle-related
variations in solar rotational splitting frequencies of low-degree
solar p modes. Although no 11-yr signals were observed, variations
on a shorter timescale $(\sim2\,\rm yrs)$ were apparent. We show
that the variations arose from complications/artifacts associated
with the realization noise in the data and the process by which the
data were analyzed. More specifically, the realization noise was
observed to have a larger effect on the rotational splittings than
accounted for by the formal uncertainties. When used to infer the
rotation profile of the Sun these variations are not important. The
outer regions of the solar interior can be constrained using
higher-degree modes. While the variations in the low-$l$ splittings
do make large differences to the inferred rotation rate of the core,
the core rotation rate is so poorly constrained, even by low-$l$
modes, that the different inferred rotation profiles still agree
within their respective $1\sigma$ uncertainties. By contrast, in
asteroseismology, only low-$l$ modes are visible and so higher-$l$
modes cannot be used to constrain the rotation profile of stars.
Furthermore, we usually only have one data set from which to measure
the observed low-$l$ splitting. In such circumstances the inferred
internal rotation rate of a main sequence star could differ
significantly from estimates of the surface rotation rate, hence
leading to spurious conclusions. Therefore, extreme care must be
taken when using only the splittings of low-$l$ modes to draw
conclusions about the average internal rotation rate of a
star.\end{abstract}


\section{Introduction}\label{section[introduction]}

The frequencies of the Sun's acoustic (p-mode) oscillations vary
throughout the solar cycle with the frequencies of the most
prominent modes being at their highest when solar activity is at its
maximum \citep[e.g.][]{Woodard1985, Palle1989, Elsworth1990,
Salabert2004, Chaplin2007, Jimenez2007}. Mid-term signals $(\lesssim
2\,\rm yr)$ can also be observed in p-mode frequencies
\citep[e.g.][]{Broomhall2009, Salabert2009, Fletcher2010}. Solar
cycle variations are also observed in other p-mode parameters such
as powers and lifetimes \citep[e.g.][]{Salabert2003, Jimenez2004a}.
With the advent of long, continuous, and high-quality asteroseismic
observations of solar-like stars \citep[e.g.][]{Chaplin2011} by the
Convection Rotation and Planetary Transits
\citep[CoRoT;][]{Baglin2006} space mission and Kepler
\citep{Borucki2010} it is possible to study changes in p-mode
properties \citep{Garcia2010, Salabert2011a} and to measure
rotational splittings \citep{Ballot2011, Beck2012}.

Solar rotation splits p modes into $2l+1$ azimuthal orders $(m)$.
The difference in frequency between azimuthal orders of a mode can
be used to infer the Sun's internal rotation profile. Variations in
the splittings of intermediate-$l$ modes $(5 \leq l \leq
300)$ have been used to observe the torsional oscillation
\citep[][and references therein]{Howe2009}. No noticeable 11-yr
solar cycle changes in the p-mode splittings of low-$l$ data have
been observed to date \citep[e.g.][]{Jimenez2002, Gelly2002,
Garcia2004, Garcia2008, Salabert2011}. \citet{Howe2000} used the
rotational splittings of medium-$l$ p modes to infer the evolution
of the rotation rate at the base of the convection zone through the
solar cycle. \citeauthor{Howe2000} found a $1.3\,\rm yr$ periodicity
in the data, however, the signal seemed to disappear after 2000
\citep{Howe2007}. Furthermore, the result was not confirmed by other
investigations \citep[e.g.][]{Basu2001, Basu2003}.
\citet{Salabert2011} observed mid-term variations in the rotational
splittings of low-$l$ modes. Here we examine whether solar
cycle-related changes, potentially including mid-term variations,
can be observed in the most up-to-date low-$l$ p-mode splittings. We
also discuss the consequences of observed variations in the
rotational splitting on solar rotation profile inversions.

Acoustic p modes are not the only oscillations that
propagate through the interior of a star. Gravity (g) modes can
propagate only in regions of stable stratification and so are
trapped within the central regions of a star, beneath the convection
zone. Mixed modes show characteristics of both p and g modes,
carrying information on both the outer envelope and the central
core. Mixed modes have been detected in stars that have evolved off
the main sequence and the splittings of mixed modes have been used
to infer the rotation rate of stellar cores \citep[][and references
therein]{Beck2012, Deheuvels2012}. However, here we consider only
the inferred average internal rotation rate of main sequence stars,
like the Sun, that do not have detectible g modes and mixed modes.
We must, therefore, rely on the splittings of low-$l$ p modes only
and so any variations in the splittings could be important.

The structure of this paper is as follows: In Section
\ref{section[splittings]} we describe how the frequency splittings
were obtained and discuss the presence of any periodicities. In
Section \ref{section[FLAG]} we compare the results obtained from the
real data with those obtained from simulated data. In Section
\ref{section[inversions]} the consequences of the variations in the splittings for inversions of the
internal rotation profile of the Sun and other stars are discussed. The main
results of the paper are summarized in Section
\ref{section[discussion]}.

\section{Extraction of rotational splitting frequencies}\label{section[splittings]}

\begin{figure*}
\centering
  \includegraphics[width=0.4\textwidth, clip]{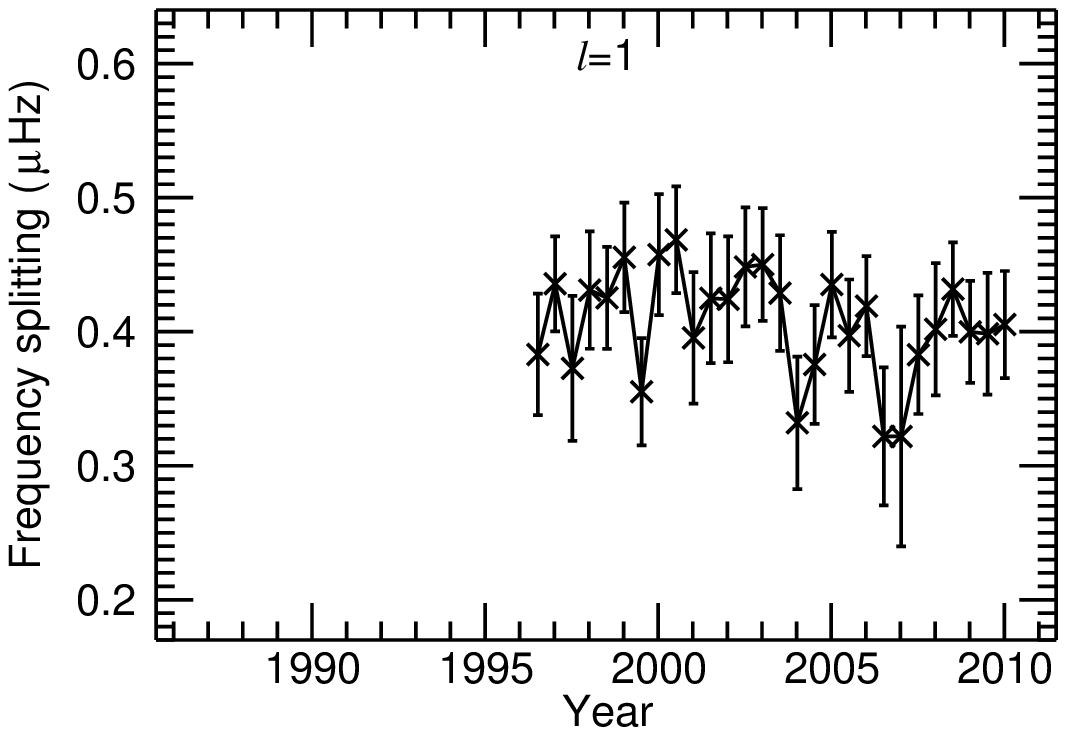}
  \includegraphics[width=0.4\textwidth, clip]{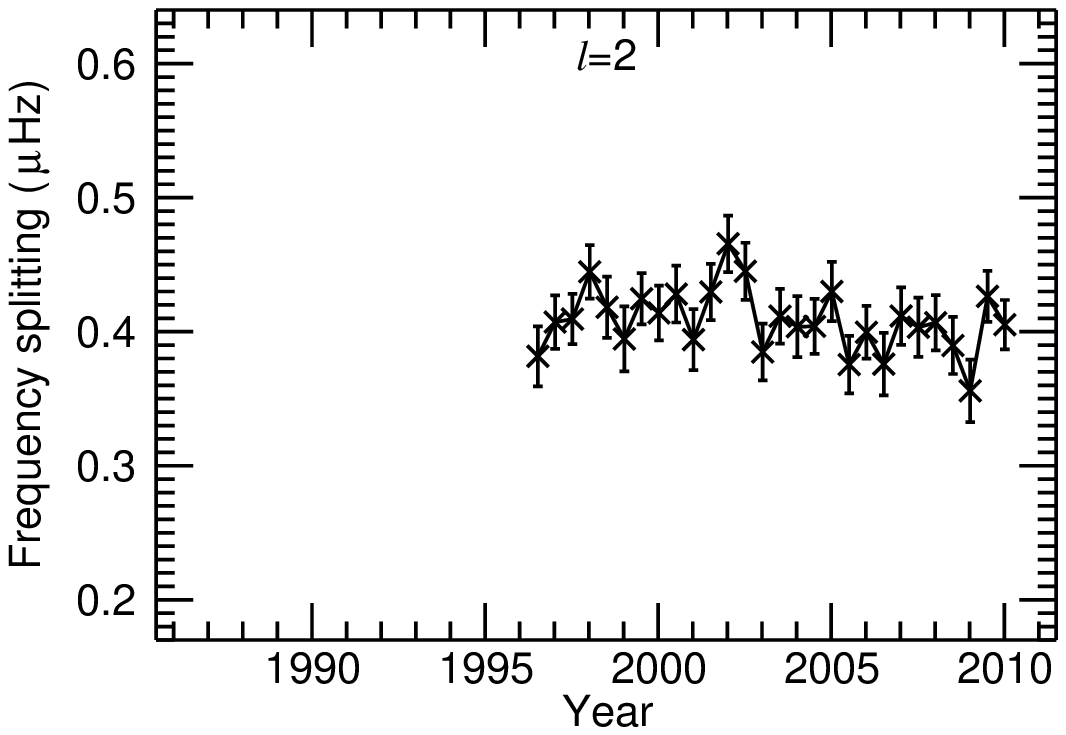}\\
  \includegraphics[width=0.4\textwidth, clip]{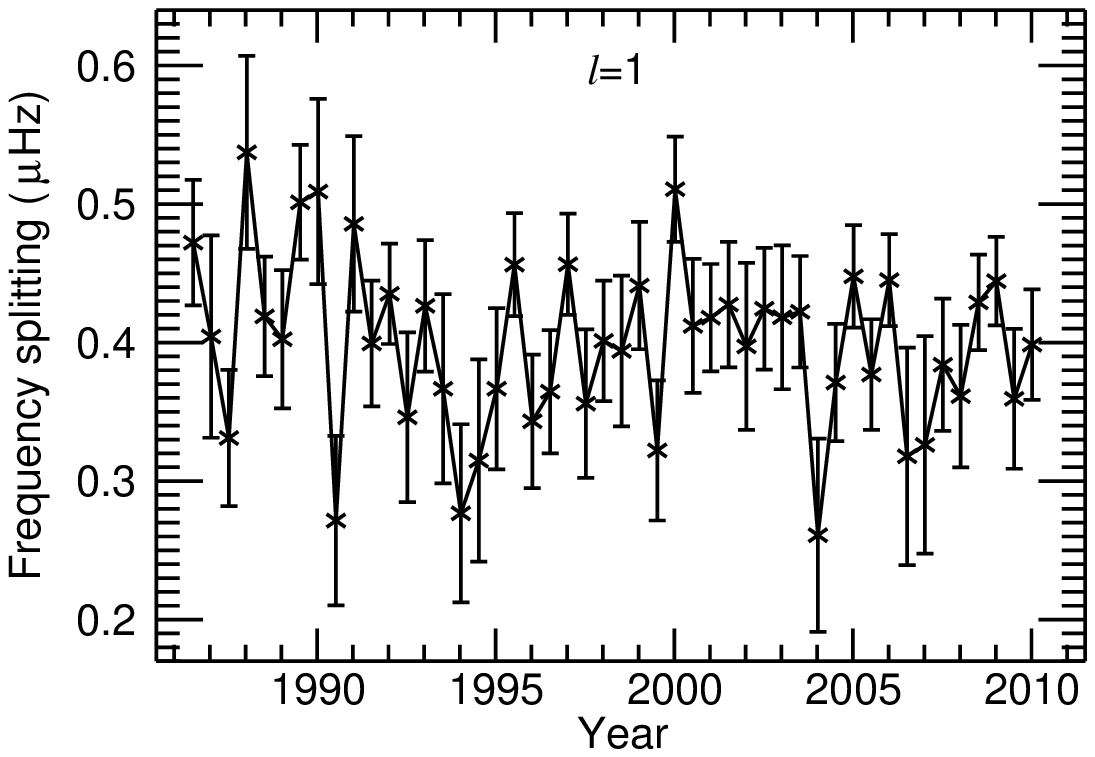}
  \includegraphics[width=0.4\textwidth, clip]{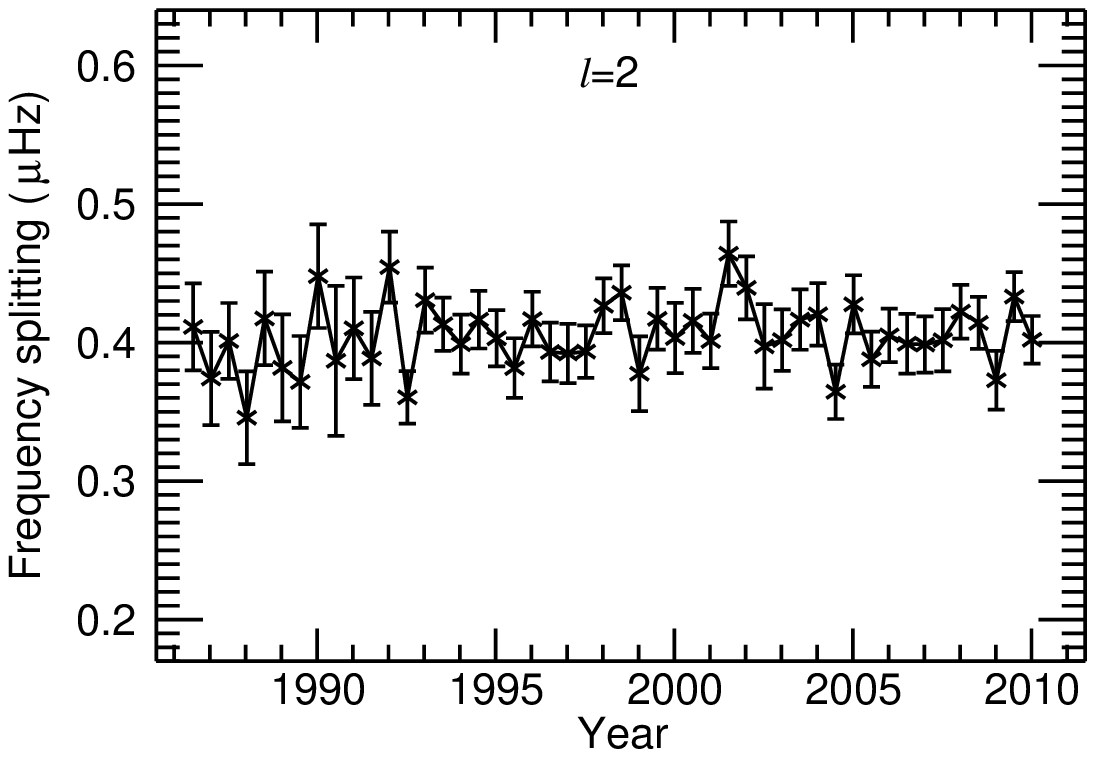}\\
  \caption{Mean splittings observed in GOLF and BiSON data. The top
  panels show the results found using the GOLF data and the bottom
  panels show the BiSON results. The left-hand panels show the $l=1$
  results and the right-hand panels show the $l=2$ results.}
  \label{figure[observed splittings]}
\end{figure*}

We have analyzed the p-mode rotational splittings observed by the
Birmingham Solar-Oscillations Network \citep[BiSON;][]{Elsworth1995,
Chaplin1996} during the last two solar cycles in their entirety i.e.
from 1986 April 14 to 2010 April 7. The Global Oscillations at Low
Frequencies \citep[GOLF;][]{Gabriel1995} instrument on board the
ESA/NASA Solar and Heliospheric Observatory (SOHO) spacecraft has
been collecting data since 1996 and so we have been able to analyze
the velocity data (following the methods described in
\citealt{Jimenez2003, Garcia2005}) covering almost the entirety of
solar cycle 23, i.e., from 1996 April 11 to 2010 April 7.

The precision with which p-mode parameters, such as the rotational
splittings, can be determined is directly related to the length of
data set under consideration. Consequently, p-mode frequencies are
often determined from data sets whose lengths are of the order of
years. However, a compromise must be made here regarding the
appropriate length of data set for study so that changes as the
solar cycle evolves can be resolved. Therefore the
observations made by GOLF and BiSON were divided into 182.5-day-long
independent subsets. After 1996 April 11, when both sets of data
were available, we ensured that the start times of the subsets from
each observational program were the same.

Estimates of the mode splittings were extracted from each subset by
fitting a modified Lorentzian model to the data using a standard
likelihood maximization method, where the rotational splitting was a
free parameter in the model.

The unweighted mean mode splitting for a particular subset was then
determined by averaging estimates in the frequency range
$2000-3100\,\rm\mu Hz$. The lower limit of this frequency
range (i.e., $2000\,\rm\mu Hz$) was the lowest frequency for which
the signal-to-noise value allowed good fits to be obtained in all
the subsets. Furthermore, below this frequency the mode frequencies
experience almost no solar cycle shift \citep[e.g.][]{Libbrecht1990,
Elsworth1994}. Above $3100\,\rm\mu Hz$ it becomes difficult to
accurately determine the mode splittings as the widths of the modes
become larger than the splitting itself and so the mode components
overlap in frequency, eventually becoming a single peak. The mean
splittings were calculated for the $l=1$ and $l=2$ modes separately.

Two different fitting codes have been used to extract the mode
frequencies \citep{Salabert2007, Fletcher2009}, both giving similar
results. For GOLF the mean absolute difference in the
estimated average splittings was $0.3\sigma$ for $l=1$ and
$0.4\sigma$ for $l=2$, and the difference in the estimated
splittings from the two codes was always within $1\sigma$. For
BiSON, the mean absolute difference in the estimated splittings was
$0.5\sigma$ for $l=1$ and $0.6\sigma$ for $l=2$. Differences in the
way the fitting procedures dealt with the BiSON window function
meant that before 1992, when the BiSON duty cycle was often below
50\,per cent, the splittings obtained from the BiSON data were
separated by up to $3\sigma$. However, after 1992 the maximum
separation of the BiSON splittings obtained by the two different
fitting methods was $1.1\sigma$. For clarity, we only show the
results of one method, which is described in
\citeauthor{Fletcher2009}.

Many authors have examined the uncertainties associated with
estimating splittings from Sun-as-a-star observations and the biases
associated with the estimated splittings are now well known
\citep[e.g.][]{Eff-Darwich1998, Appourchaux2000, Chaplin2006,
Garcia2008}. We follow the advice of \citet{Chaplin2006} in order to
minimize these biases. For example, \citeauthor{Chaplin2006} found
that one of the main sources of bias in estimated splittings arises
if the amplitude ratios of the $m$ components inside a multiplet are
not set correctly. We therefore fix these ratios at previously
determined and accepted values for BiSON and GOLF
\citep{Chaplin2006, Salabert2011b}. Furthermore, the amplitude of
the majority of known systematic errors become appreciable above
approximately $3400\,\rm\mu Hz$, which is above the upper limit on
the frequency range we examine here.

\subsection{Variation with time of the rotational splitting frequencies}
Let us now consider whether the rotational splitting measured with
the methodology outlined above show any significant variations with
time. Fig. \ref{figure[observed splittings]} shows the mean
determined splittings as a function of time, $a_l(t)$, for GOLF and
BiSON. There is no obvious evidence for an 11-yr solar cycle effect
on the splittings. This is consistent with previous results
\citep[e.g.][]{Jimenez2002, Gelly2002, Garcia2004, Garcia2008,
Salabert2011}. However, there are discernible mid-term $(\sim2\,\rm
yr)$ variations, which is also in agreement with the results of
\citet{Salabert2011}. There is significant correlation between the
variations in the splittings observed in the two data sets (see
Table \ref{table[correlations]}), such that there is less than a
0.05\,per cent probability of the correlations occurring by chance.

\begin{table}
\centering \caption{Pearson's and Spearman's rank correlations between the BiSON and GOLF splittings}\label{table[correlations]}
\begin{tabular}{ccc}
  \hline
  \noalign{\smallskip}
   & $l=1$ & $l=2$ \\
  \noalign{\smallskip}
  \hline
  \noalign{\smallskip}
  Pearson's & $0.84$ & $0.63$ \\
  Spearman's & $0.80$ & $0.63$ \\
  \noalign{\smallskip}
  \hline
\end{tabular}
\end{table}

\begin{figure}
  \centering
  \includegraphics[width=0.4\textwidth, clip]{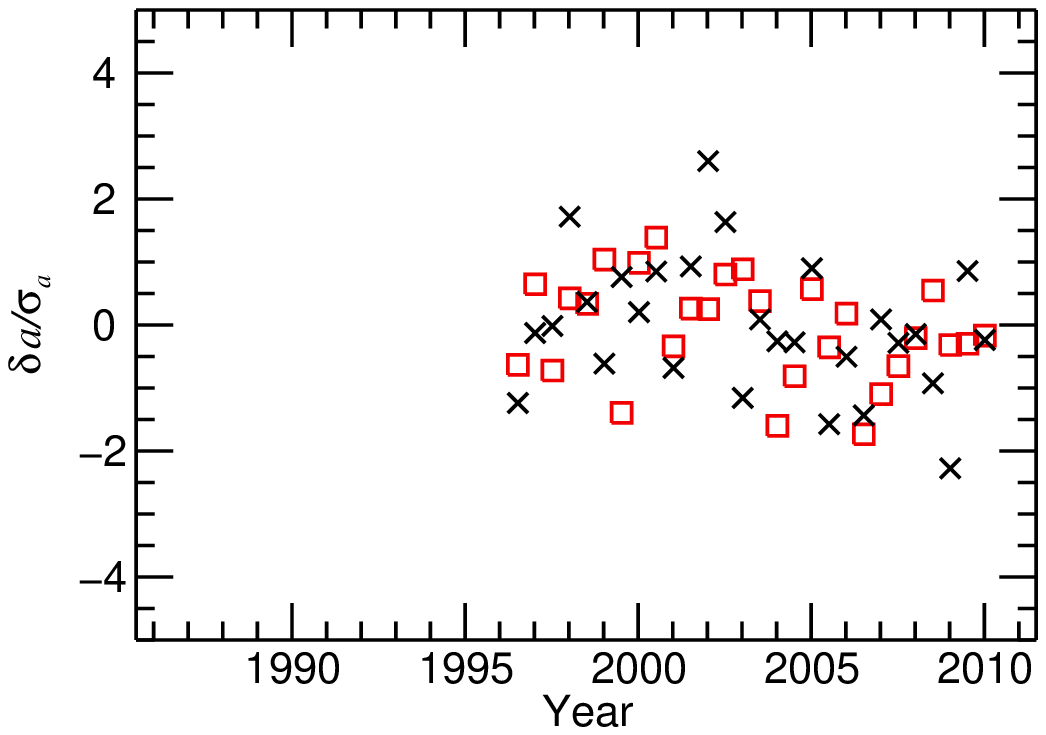}\\
  \includegraphics[width=0.4\textwidth, clip]{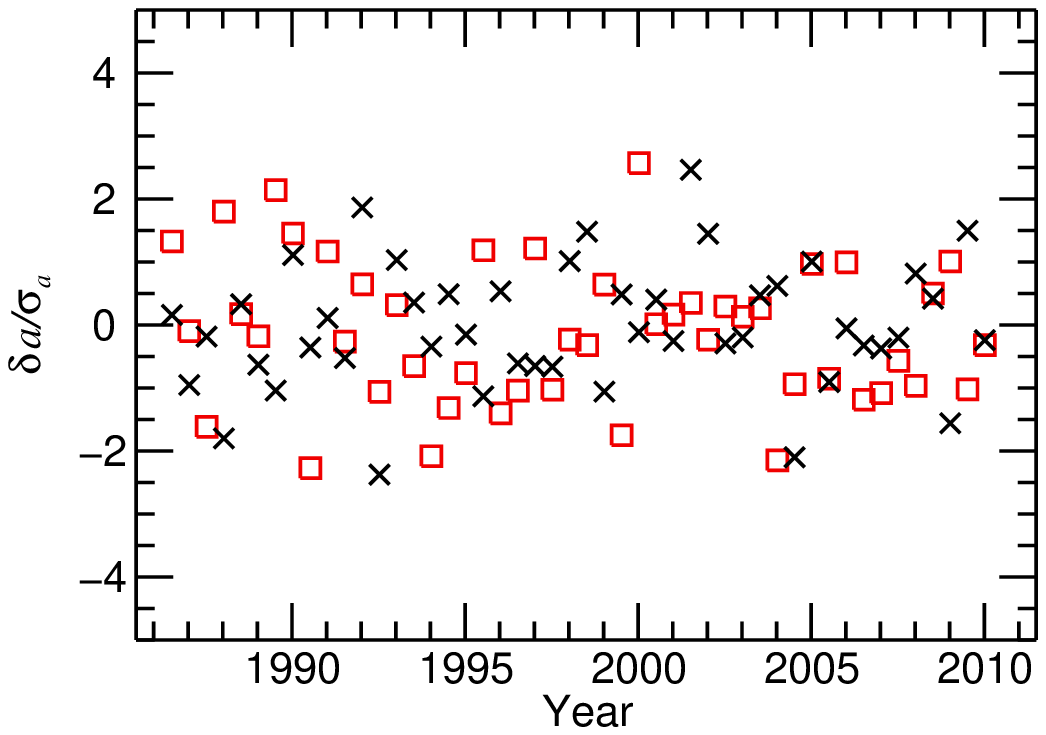}\\
  \caption{Relative variation in the error normalized residuals, $(\delta a_l(t)/\sigma_a)$. The top panel shows the results for GOLF and the bottom panel shows the results for BiSON. The red squares show the $l=1$ results and the black crosses show the $l=2$ results.}\label{figure[splitting variation]}
\end{figure}

There is no correlation between the $l=1$ and $l=2$ mode splittings
for either the GOLF or the BiSON data. The splittings of the $l=1$
modes show more variation than the $l=2$ mode splittings. This is
not surprising and indicates that, in Sun-as-a-star data, the
rotational splittings estimated from $l=2$ modes are more stable
than the splittings estimated using $l=1$ modes \citep[see
e.g.][]{Chaplin2006}. However, as we now show, this difference is
accounted for by the formal uncertainties associated with the
estimated splittings. We define $\bar{a_l}$ to be the temporal mean
of $a_l(t)$ and $\delta a_l(t)=a_l(t)-\bar{a_l}$. Fig.
\ref{figure[splitting variation]}, which shows $\delta
a_l(t)/\sigma_a$ (where $\sigma_a$ is the formal uncertainty
associated with $a_l(t)$), demonstrates that the scatter in the
plots is similar for $l=1$ and 2, indicating that the relative
variation in the splittings is comparable. Table \ref{table[sd
variations]} gives the standard deviations of the error normalized
residuals, $\delta a_l(t)/\sigma_a$. The standard deviations of the
normalized residuals are close to unity for both GOLF and BiSON
data, which is what we would expect to observe for a totally random
process with appropriate error bars.

\begin{table}
\centering \caption{Standard deviation of the error normalized residuals (plotted in Fig. \ref{figure[splitting variation]}).}\label{table[sd variations]}
\begin{tabular}{ccc}
  \hline
  \noalign{\smallskip}
  $l$ & GOLF & BiSON\\
  \noalign{\smallskip}
  \hline
  \noalign{\smallskip}
  1 & $0.82$ & $1.14$ \\
  2 & $1.07$ & $1.00$ \\
  \noalign{\smallskip}
  \hline
\end{tabular}
\end{table}

\subsection{Periodograms of the observed splittings}
To further investigate whether the observed variations are
significant, we computed periodograms of the observed splittings.
These periodograms are plotted in Fig. \ref{figure[observed
periodograms]}. In calculating the periodograms the data have been
oversampled by a factor of 10. The cut-off frequency at the low end
of the periodograms, below which no information can be obtained, is
$0.07\,\rm yr^{-1}$ and $0.042\,\rm yr^{-1}$ for GOLF and BiSON
respectively. Also plotted in Fig. \ref{figure[observed
periodograms]} are the 1\,per cent false alarm significance levels
\citep{Chaplin2002}, which were determined using Monte Carlo
simulations. 200,000 noise data sets were simulated, using a normal
distribution random number generator, to mimic those plotted in Fig.
\ref{figure[observed splittings]}. The standard deviation of each
point on the data set was taken as the $1\sigma$ uncertainty
associated with the rotational splittings plotted in Fig.
\ref{figure[observed splittings]}. The simulated data sets were then
used to create periodograms and the amplitudes observed in the
simulated periodograms were used to define the 1\,per cent false
alarm significance levels. The 1\,per cent significance level is
somewhat arbitrary, but was chosen a priori so that the number of
expected false detections is less than unity \cite[see
e.g.][]{Chaplin2002, Broomhall2010}.

\begin{table}
\centering \caption{Null probability of observing the peaks in the
BiSON periodogram (Fig. \ref{figure[observed periodograms]}) in
noise.}\label{table[BiSON probabilities]}
\begin{tabular}{ccc}
  \hline
  \noalign{\smallskip}
  $l$ & Frequency $\rm(yr^{-1})$ & Probability (\%)\\
  \noalign{\smallskip}
  \hline
  \noalign{\smallskip}
  1 & $0.08$ & $0.190$ \\
  1 & $0.58$ & $0.797$ \\
  1 & $0.67$ & $0.679$ \\
  2 & $0.92$ & $0.337$ \\
  \noalign{\smallskip}
  \hline
\end{tabular}
\end{table}

There are no significant peaks (at a 1 per cent level) in the
periodogram of the GOLF data indicating that there are no
significant periodicities in the GOLF rotational splittings.
However, there are three significant peaks in the $l=1$ BiSON
periodogram and one significant periodicity in the BiSON $l=2$
splittings. The significant peak in the BiSON $l=2$ periodogram (at
$\sim0.92\,\rm yr^{-1}$) does not correspond to any of the
significant peaks in the BiSON $l=1$ periodogram.

Table \ref{table[BiSON probabilities]} contains the
probabilities of those peaks in the BiSON periodogram where the
probability of observing the peaks by change is less than 1\,per
cent. Although two of the $l=1$ peaks are only marginally
significant the other $l=1$ peak (at $0.67\,\rm yr^{-1}$) and the
$l=2$ peak are well below the significance level. The uncertainties
associated with the splittings would have to be increased by 10\,per
cent for the $l=2$ modes and 17\,per cent for the $l=1$ modes before
these peaks are no longer significant.

\begin{figure}
\centering
  \includegraphics[width=0.4\textwidth, clip]{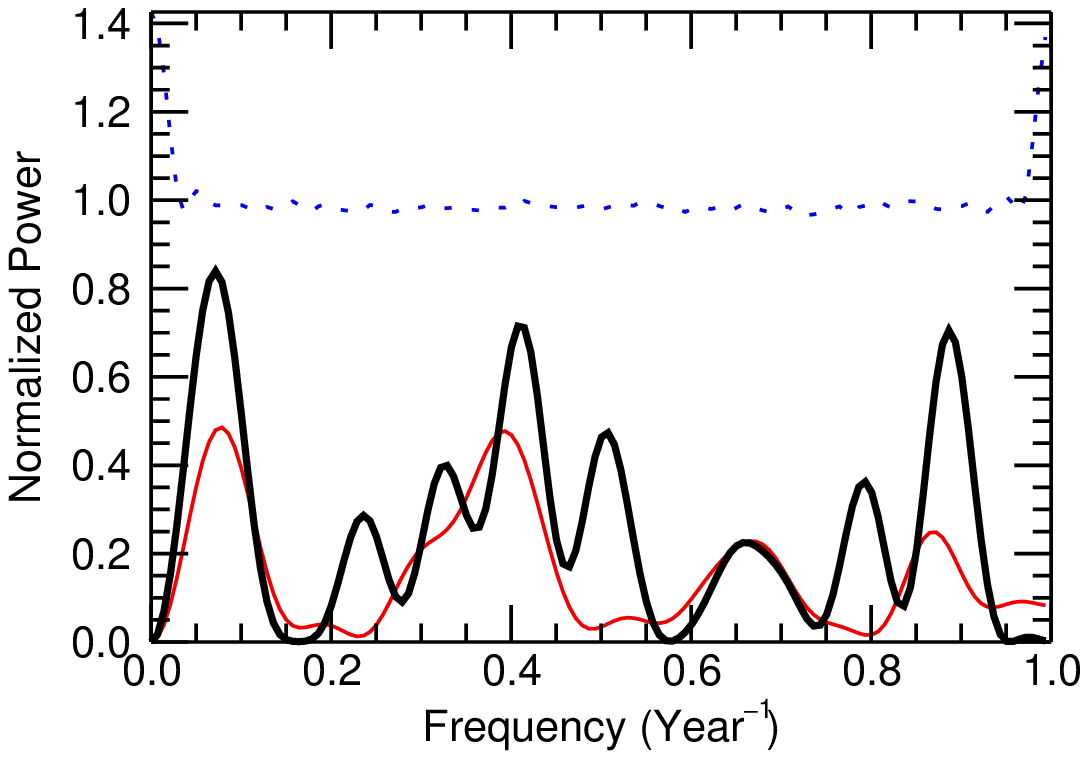}\\
  \includegraphics[width=0.4\textwidth, clip]{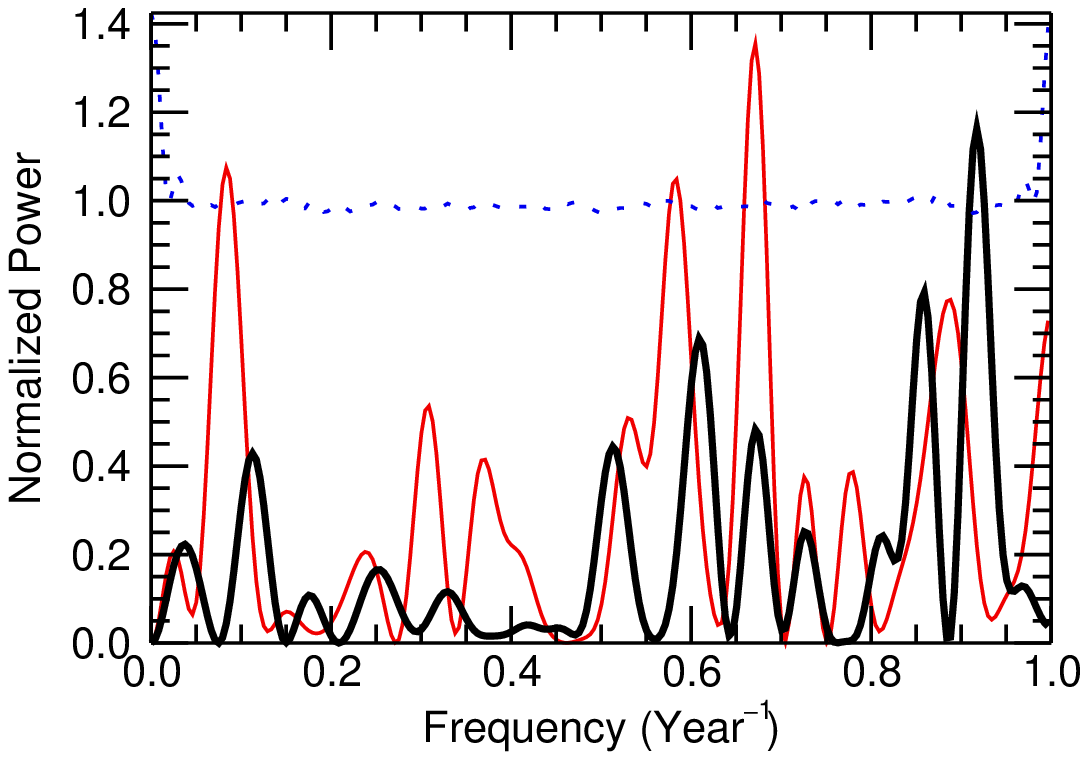}\\
  \caption{Periodograms of the mean splittings observed in GOLF
  (top panel) and BiSON (bottom panel) data. Each periodogram has been normalized so
  that the power is unity when the mean significance level is 1\,per cent.
  This means that the $l=1$ and 2 significance levels are now coincident.
  In each panel the red, thin line represents the $l=1$ results,
  the black, thick line represents the $l=2$ results and the blue
  dotted line represent the 1\,per cent significance level.}
  \label{figure[observed periodograms]}
\end{figure}

\section{Comparison with solarFLAG data}\label{section[FLAG]}

\begin{figure}
\centering
  \includegraphics[width=0.4\textwidth, clip]{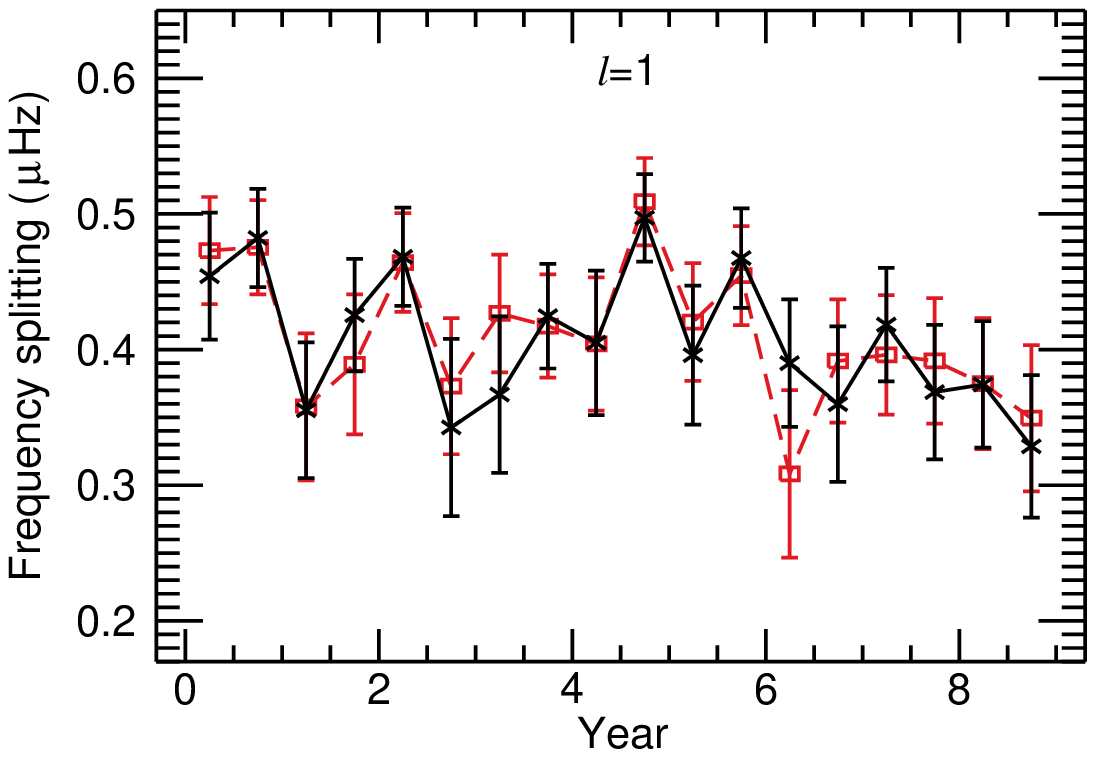}\\
  \includegraphics[width=0.4\textwidth, clip]{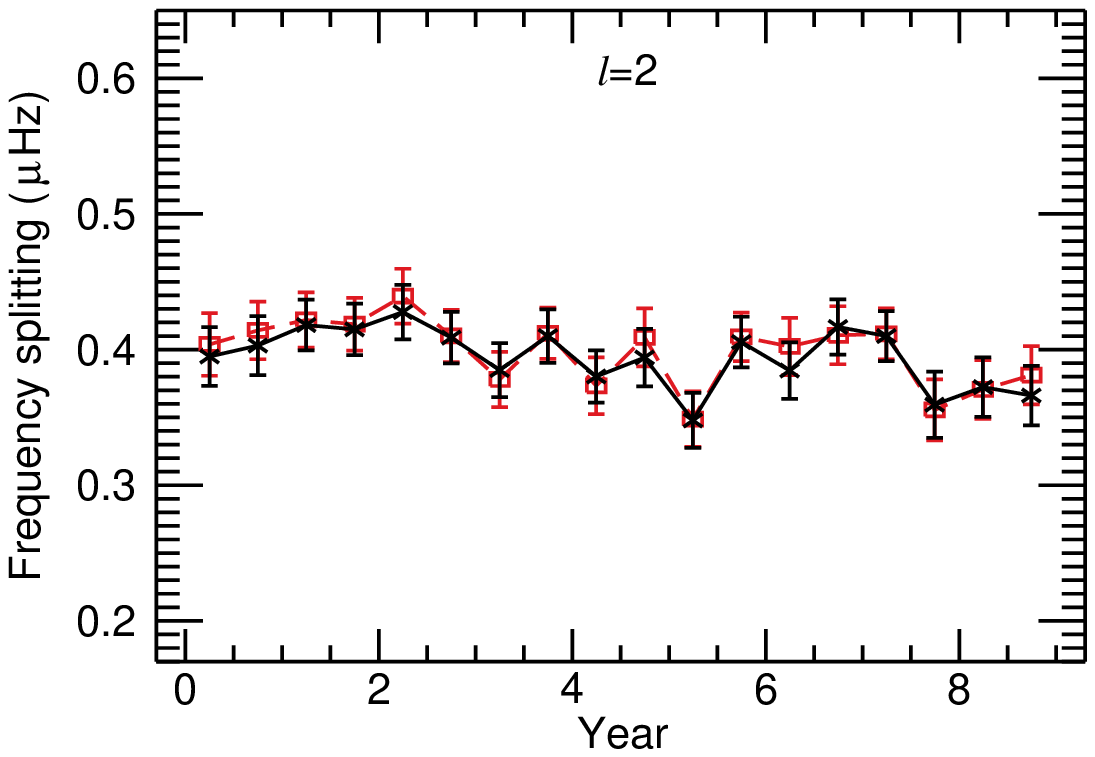}\\
  \caption{Mean splittings observed in one of the simulated solarFLAG data sets (FLAGa). The top
  panel shows the $l=1$ results and the bottom panel shows the
  $l=2$ results. The black crosses with the solid line represent the
  results when the solarFLAG data had a duty cycle of 100\,per cent. The red squares with
  the dashed line represent the results when the 182.5-d-long solarFLAG subsets were
  given a BiSON window function with a
  duty cycle of 88\,per cent.}
  \label{figure[FLAG splittings]}
\end{figure}

To help determine whether the signals observed in the BiSON data are
really solar in origin (e.g. perhaps associated with the solar
cycle) or simply an artifact of the data analysis procedure we
examined data that were simulated for the solar Fitting at Low
Angular degree Group \citep[solarFLAG;][]{Chaplin2006}. The data
were simulated in the time domain and were designed to mimic
Sun-as-a-star observations. The simulated oscillations were
stochastically excited and damped with lifetimes analogous to those
of real solar oscillations. Five sets of solarFLAG data were
examined. The input mode frequencies were the same in each simulated
data set and were constant in time. The mode splittings were fixed
at $0.4\,\rm\mu Hz$. However, the data simulated the stochastic
nature of the mode excitation giving access to different mode and
noise realizations. The simulated data were approximately 9\,yr in
length but were split into subsets of 182.5\,d for analysis.

The splittings obtained from a typical solarFLAG data set (FLAGa)
are plotted in Fig. \ref{figure[FLAG splittings]}. Almost as much
variation is observed in the splittings obtained from the solarFLAG
data as was observed in the real data. Table \ref{table[abs
deviation]} shows the mean observed splittings and the maximum
absolute deviation from the mean for both the real and simulated
data. The maximum absolute deviations of the BiSON data are slightly
larger than all of the simulated solarFLAG data sets, while the GOLF
data are in reasonable agreement with the solarFLAG data. Since the
only source of variation in the solarFLAG data is the realization
noise these results imply that the changes seen in the real data are
more than likely due to realization noise.

\subsection{Impact of the BiSON window function}
The discrepancy between the maximum absolute deviations observed in
the BiSON results and the GOLF and solarFLAG results can be
explained by the impact of the window functions of the respective
data sets. The duty cycles of the BiSON subsets range from 22 to
88\,per cent, with the average duty cycle being 69\,per cent.
Normally the duty cycle of the GOLF data is above 90\,per cent and
the mean duty cycle is 95\,per cent. However, the lowest observed
duty cycle of the GOLF subsets was 42\,per cent, which coincides
with when control of SOHO was temporarily lost, known as the ``SOHO
vacation''. BiSON window functions were applied to the GOLF data
such that the 182.5\,d BiSON subset from which the window function
was extracted matched in time the GOLF subset to which it was
applied. Table \ref{table[abs deviation]} shows that this increases
the maximum absolute deviations of the GOLF data. Furthermore,
although not shown here, a marginally significant peak (at a 1\,per
cent significance level) was observed at $\sim0.9\,\rm yr^{-1}$ in
the periodogram of the $l=2$ splittings.

Various BiSON window functions were also imposed on the
solarFLAG data, with duty cycles covering the range observed here.
In addition to the splittings observed in the 100 per cent duty
cycle solarFLAG data, Fig. \ref{figure[FLAG splittings]} also shows
the splittings obtained from one set of solarFLAG data when a
182.5-d-long BiSON window function, with a duty cycle of
88\,per cent, was imposed upon each 182.5-d-long subset. This is
just one example of the window functions that were imposed. The same
BiSON window function was imposed on each FLAG subset so that the
average duty cycle could be varied systematically. When some window
functions were imposed the maximum absolute deviations were observed
to increase to values similar to those observed in the BiSON data.
However, in other cases the maximum absolute deviation was observed
to decrease. Table \ref{table[fill abs deviation]} shows that,
although altering the window function changes the maximum absolute
deviation, the obtained deviation is not dependent on the value of
the duty cycle. For example, the $l=1$ rotational splittings
observed in FLAGa varied more when the duty cycle was 88\,per cent
than when the duty cycle was 47\,per cent.

This is consistent with the surmise that the observed variations are
due to realization noise and that the window function could affect
the variations in the rotational splittings both constructively and
destructively via its interaction with the noise.
Consider a period of time when the realization noise shifts the
obtained splitting away from the true value. If the window function
blanks this period of time (i.e. there is no data available) then the estimated rotational
splitting will be close to the true value and the observed variation
in the rotational splittings will decrease. However, the window
function could equally favour times when the realization noise
shifts the observed rotational splitting away from the true value.
The variation in the rotational splittings will then increase.

\begin{figure*}
\centering
  \includegraphics[width=0.3\textwidth, clip]{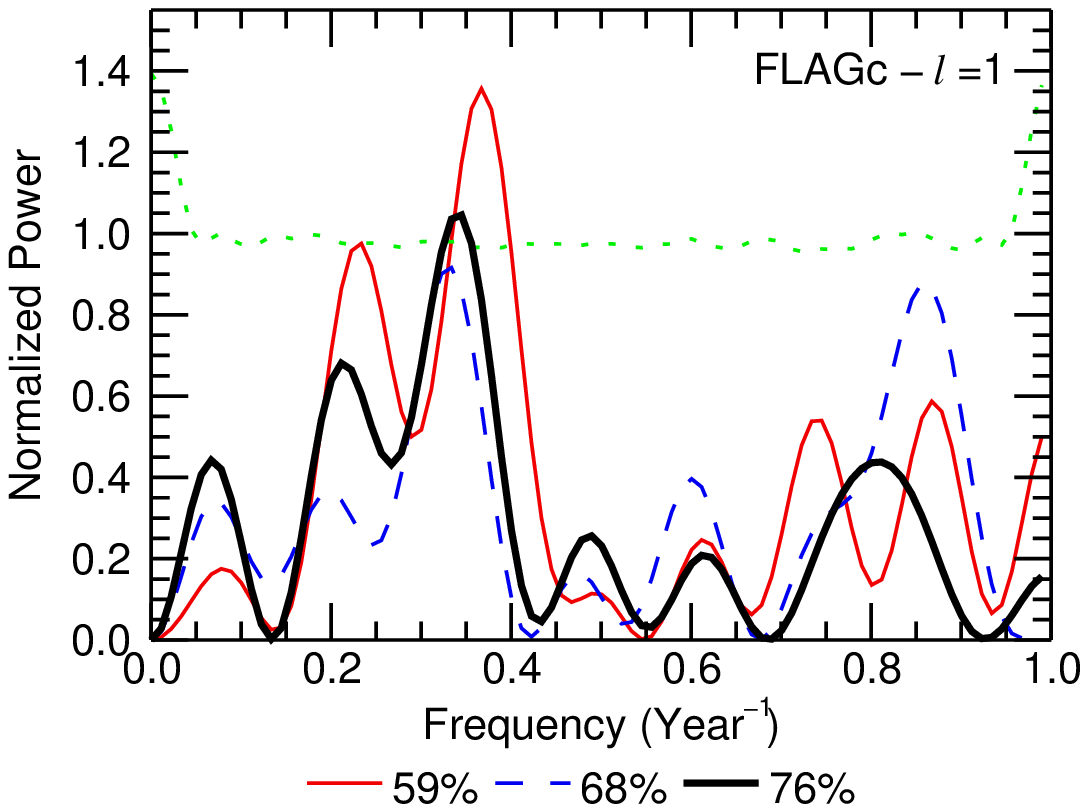}
  \includegraphics[width=0.3\textwidth, clip]{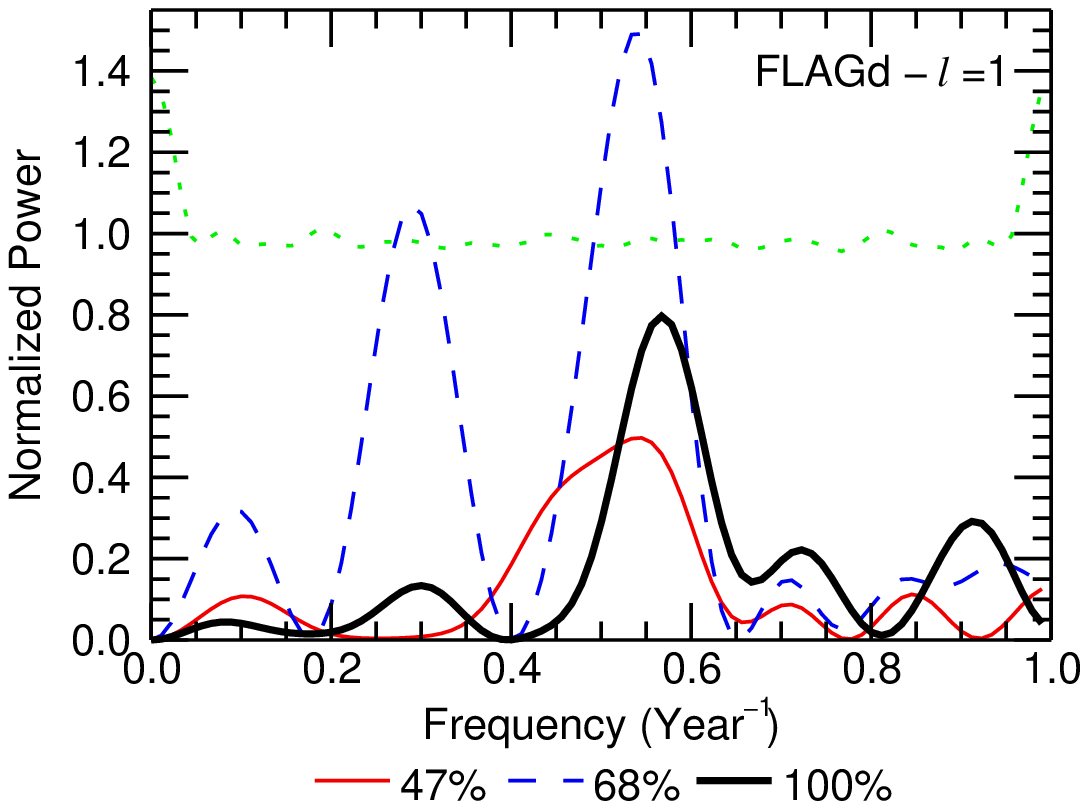}
  \includegraphics[width=0.3\textwidth, clip]{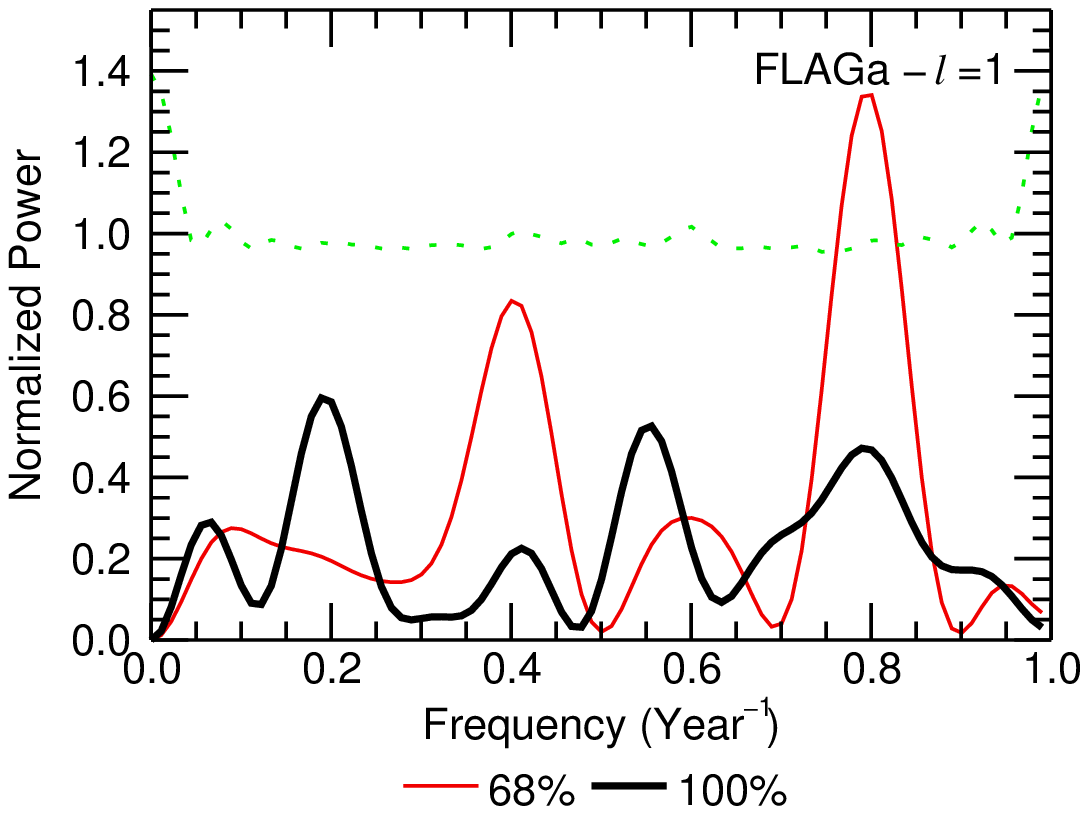}\\
  \includegraphics[width=0.3\textwidth, clip]{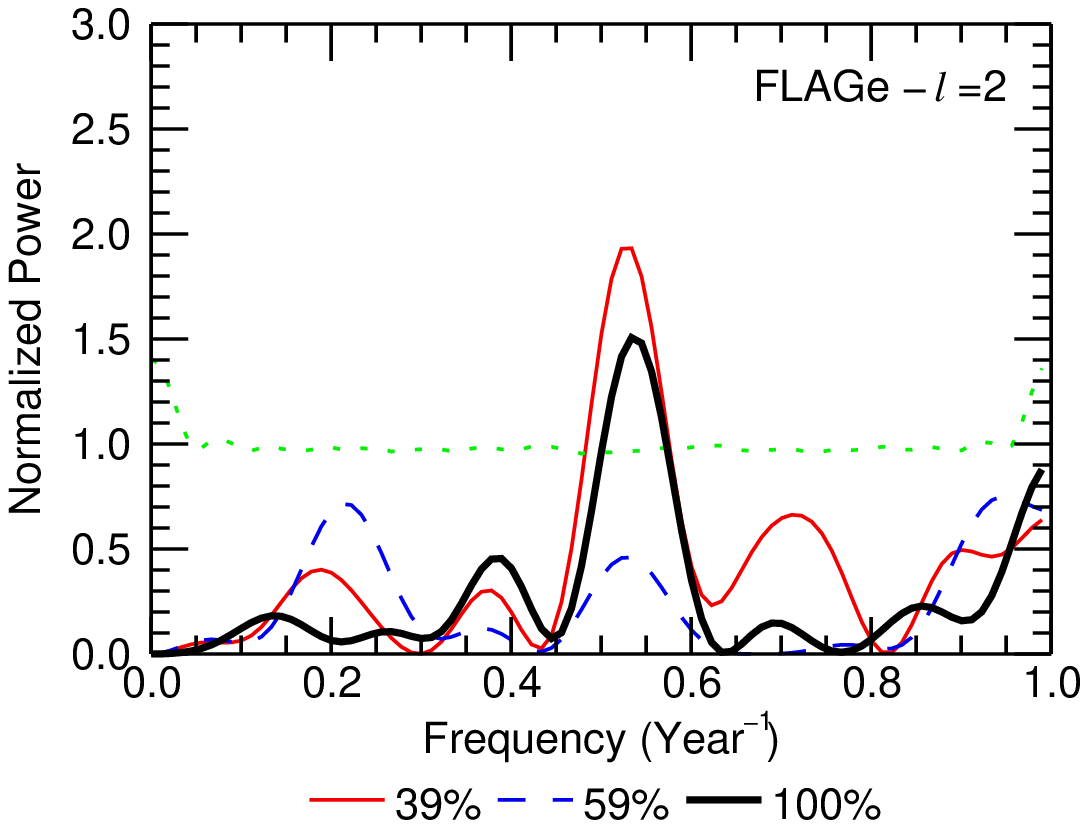}
  \includegraphics[width=0.3\textwidth, clip]{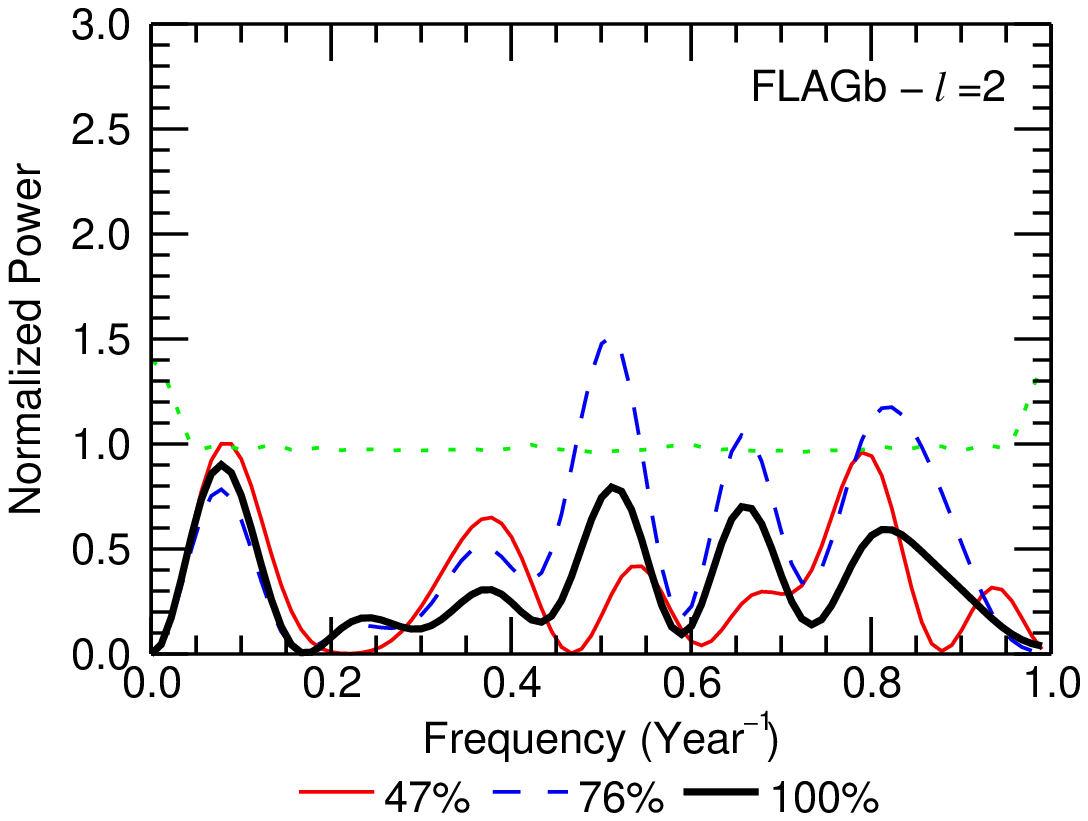}
  \includegraphics[width=0.3\textwidth, clip]{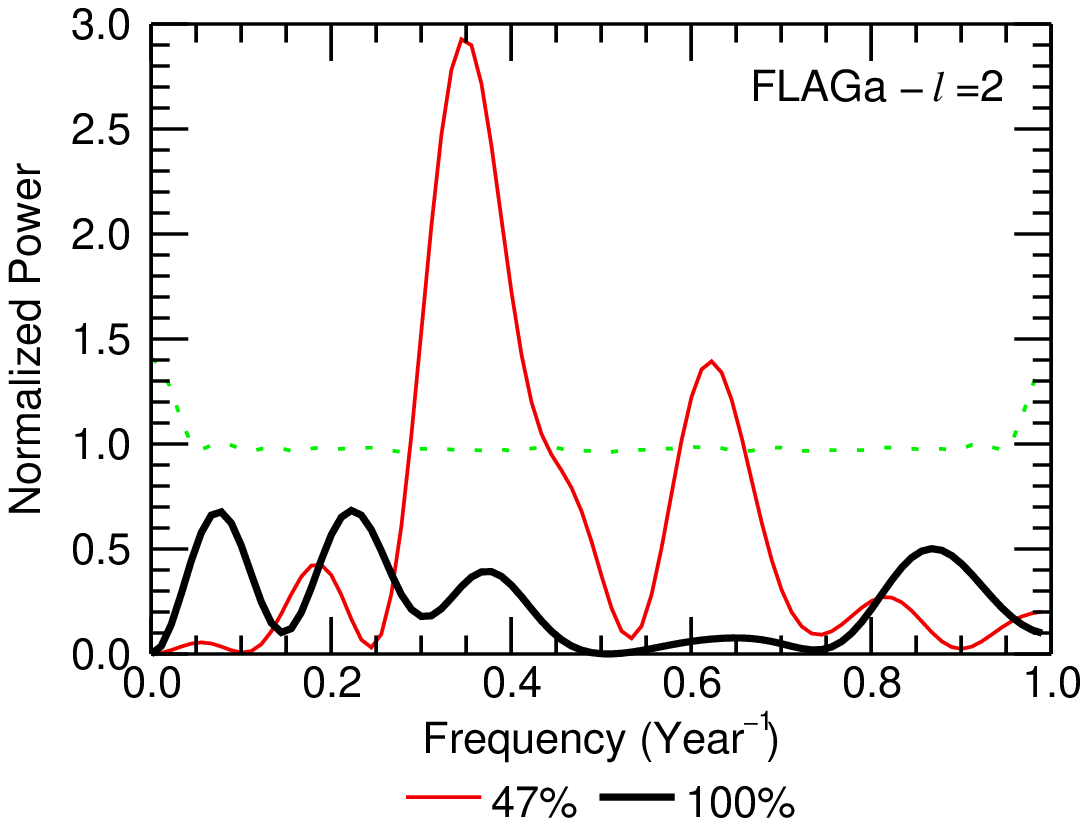}\\
  \caption{Periodograms of the mean splittings observed in a selection of the
  simulated solarFLAG data sets. The periodograms have been normalized so
  that the mean 1\,per cent significance level (green dotted lines) is unity. The duty cycle of the BiSON window functions
  imposed on each subset are
  indicated in the legends. The top row shows $l=1$ results, while the
  bottom row shows $l=2$ results. The left and middle columns show
  that the power of peaks in the periodograms can both increase and
  decrease as the duty cycle decreases. The right-hand column shows
  that changes in the duty cycle can give rise to very different
  periodograms. }
  \label{figure[FLAG periodograms]}
\end{figure*}

This hypothesis is strengthened further by analysis of the
periodograms of the solarFLAG data. Fig. \ref{figure[FLAG
periodograms]} shows periodograms of the splittings obtained from
the solarFLAG data sets when various BiSON window functions
were imposed. We start by considering the top-left panel of
Fig. \ref{figure[FLAG periodograms]}, which shows periodograms of
the $l=1$ splittings obtained from the FLAGc data when BiSON window
functions were imposed such that the duty cycle of the data was
59\,per cent, 68\,per cent and 76\,per cent respectively. Even
though the splittings included in the simulated FLAG data were
constant, significant peaks (at a 1\,per cent false alarm level) are
observed in the periodograms. Significant peaks are observed when
the duty cycle was both 59\,per cent and 76\,per cent but not
68\,per cent. Similarly in the top middle panel of Fig.
\ref{figure[FLAG periodograms]} significant peaks (at a 1\,per cent
false alarm level) are observed in periodograms of the $l=1$
splittings obtained from the FLAGd data when the duty cycle is
68\,per cent but not 47\,per cent or 100\,per cent. This indicates
that the likelihood of observing a significant peak is not simply a
function of the duty cycle. The bottom left and middle panels show
that the same is true for the $l=2$ splittings: The bottom left
panel shows that significant peaks are observed in the $l=2$
splittings obtained from the FLAGe data when the duty cycle was
39\,per cent and 100\,per cent but not 59\,per cent. The bottom
middle panel shows that significant peaks are observed in the $l=2$
splittings obtained from the FLAGb data when the duty cycle was
76\,per cent but not 47\,per cent or 100\,per cent. This shows that
more important than the actual duty cycle itself is the manner in
which the window function interacts with the realisation noise. The
top and bottom right-hand panels of Fig. \ref{figure[FLAG
periodograms]} show just how different the periodograms can look
simply by changing the window function of the data.

\begin{table*}
\centering \caption{Mean splittings and maximum absolute deviation
from the mean.}\label{table[abs deviation]}
\begin{tabular}{ccccc}
  \hline
  \noalign{\smallskip}
  Data set & \multicolumn{2}{c}{Mean splitting $(\rm\mu Hz)$} &
  \multicolumn{2}{c}{Maximum absolute deviation $(\rm\mu Hz)$} \\
  & $l=1$ & $l=2$ & $l=1$ & $l=2$ \\
  \noalign{\smallskip}
  \hline
  \noalign{\smallskip}
  BiSON & $0.411\pm0.007$ & $0.406\pm0.003$ & $0.150\pm0.070$ &
   $0.061\pm0.034$ \\
  GOLF & $0.412\pm0.008$ & $0.410\pm0.004$ & $0.090\pm0.082$ &
  $0.056\pm0.021$ \\
  GOLF (BiSON window function)& $0.414\pm0.008$ & $0.412\pm0.004$ & $0.125\pm0.073$ &
  $0.068\pm0.038$ \\
  \noalign{\smallskip}
  \hline
  \noalign{\smallskip}
  FLAGa & $0.424\pm0.010$ & $0.396\pm0.005$ & $0.095\pm0.054$ &
  $0.048\pm0.021$ \\
  FLAGb & $0.423\pm0.010$ & $0.397\pm0.005$ & $0.064\pm0.053$ &
  $0.059\pm0.019$ \\
  FLAGc & $0.426\pm0.010$ & $0.406\pm0.005$ & $0.096\pm0.076$ &
  $0.035\pm0.024$ \\
  FLAGd & $0.421\pm0.011$ & $0.401\pm0.005$ & $0.097\pm0.077$ &
  $0.042\pm0.045$ \\
  FLAGe & $0.413\pm0.010$ & $0.408\pm0.005$ & $0.069\pm0.068$ &
  $0.054\pm0.022$\\
  \noalign{\smallskip}
  \hline
\end{tabular}
\end{table*}

\begin{table*}
\centering \caption{Mean splittings and maximum absolute deviation
from the mean estimated from the FLAGa data set when different
window functions were imposed on the data.}\label{table[fill abs
deviation]}
\begin{tabular}{ccccc}
  \hline
  \noalign{\smallskip}
  Duty cycle (\%) & \multicolumn{2}{c}{Mean splitting $(\rm\mu Hz)$} &
  \multicolumn{2}{c}{Maximum absolute deviation $(\rm\mu Hz)$} \\
  & $l=1$ & $l=2$ & $l=1$ & $l=2$ \\
  \noalign{\smallskip}
  \hline
  \noalign{\smallskip}
  100 & $0.424\pm0.010$ & $0.396\pm0.005$ & $0.095\pm0.054$ &
  $0.048\pm0.021$ \\
  88 & $0.426\pm0.010$ & $0.400\pm0.005$ & $0.117\pm0.063$ &
  $0.051\pm0.021$ \\
  76 & $0.416\pm0.010$ & $0.396\pm0.005$ & $0.110\pm0.054$ &
  $0.043\pm0.023$ \\
  68 & $0.435\pm0.010$ & $0.392\pm0.005$ & $0.121\pm0.047$ &
  $0.055\pm0.024$ \\
  59 & $0.417\pm0.009$ & $0.382\pm0.005$ & $0.103\pm0.062$ &
  $0.046\pm0.020$\\
  47 & $0.421\pm0.008$ & $0.399\pm0.005$ & $0.095\pm0.056$ &
  $0.088\pm0.023$\\
  38 & $0.414\pm0.010$ & $0.379\pm0.006$ & $0.130\pm0.063$ &
  $0.053\pm0.035$\\
  22 & $0.406\pm0.012$ & $0.403\pm0.008$ & $0.162\pm0.078$ &
  $0.138\pm0.029$\\
  \noalign{\smallskip}
  \hline
\end{tabular}
\end{table*}

We note here that Tables \ref{table[abs deviation]} and \ref{table[fill abs deviation]} indicate that the mean $l=1$ rotational splittings are
consistently higher than the input value of $0.4\,\rm\mu Hz$, because the rotationally split components overlap in frequency giving rise to a positive bias \citep{Chaplin2006}.

The results of the solar-FLAG simulations show that
significant quasi-periodic variations can be observed in the
splittings even when no actual, underlying variations in the
splittings are present. The BiSON window function can act to both
enhance or reduce the significance of the variation.

\section{Consequences for rotation profile inversions}\label{section[inversions]}

Let us examine the consequences of these apparent variations in the
splittings for both helioseismic rotation inversion profiles and
asteroseismic inferences of the mean internal rotation rate. Each
rotational splitting is a spatially weighted average of certain
volumes of a star's internal rotation rate and various inversion
techniques have been developed to infer the Sun's internal rotation
profile from the observed splittings. We now consider the
consequences of the observed variations in the splittings on
inferences that can be made about the rotation profiles of the Sun
and other stars.

\subsection{Inversions of the Sun's internal rotation profile}

We have seen that, although the mean BiSON $l=1$ splitting is
$0.411\pm0.007\,\rm\mu Hz$, the maximum and minimum observed
splittings are $0.537\pm0.070\,\rm\mu Hz$ and $0.261\pm0.070\,\rm\mu
Hz$ respectively. We have already shown that the observed variation
in the splittings is just an artifact associated with the
realization noise in the data and the process by which the data were
analyzed. However, one may erroneously ascribe the splitting
variations to a physical change in the rotation rate of a region of
the solar interior. We have split the solar interior into 4 regions;
the solar core $(0.00R_{\odot}\le r<0.20R_{\odot})$, the radiative
zone $(0.20R_{\odot}\le r<0.70R_{\odot})$, the convection zone
$(0.70R_{\odot}\le r<0.95R_{\odot})$, and the near-surface shear
layer $(0.95R_{\odot}\le r<1.00R_{\odot})$. We have then determined
by how much the rotation rate of that layer would need to change,
relative to an average 2D Regularized Least Squares (RLS) inversion
\citep{Schou1994} of the solar rotation profile, to explain the
observed differences in the rotational splittings. The inversions
\citep{Thompson1996, Howe2005} were made using 15\,yrs of Global
Oscillations Network Group \citep[GONG,][]{Harvey1996} data. This
analysis is similar to that performed for the Sun by
\citet{Elsworth1995a} and more details can be found in Appendix
\ref{section[appendix inversions forward]}.

Using the above mentioned maximum and minimum measured splittings we
find that the maximum departure from the mean BiSON splitting was
$0.150\pm0.070\,\rm\mu Hz$. Table \ref{table[splitting inversions]}
shows the change in rotation rate, $\textrm{d}(\Omega/2\pi)$,
required to change the observed splitting of $l=1$ and 2, $n=17$
modes by $0.15\,\rm\mu Hz$ (results for other modes can be found in
Appendix \ref{section[appendix inversions forward]}). The results
indicate that the changes required in the rotation rate of
near-surface regions to explain a difference in the splitting of
$0.15\,\rm\mu Hz$ are not unreasonably large. However, observations
of higher-$l$ modes rule out variations in the near-surface regions
even of this magnitude (see Appendix \ref{section[appendix
inversions]}). Information from higher-$l$ modes dominate and
constrain inversions of the rotation in outer regions of the solar
interior at better than a 0.5\,per cent level. Although variations
with time in the higher-$l$ splittings are detectable (due to the
torsional oscillation) these variations are less than $0.001\,\rm\mu
Hz$ \citep[e.g.][]{Howe1999} and so they are approximately 100 times
smaller than the deviations observed here. Large changes in the core
rotation rate are required to explain the observed change in the
rotational splittings. Although the required changes are so large as
to be unlikely\footnote{The rotation rate in the core is not well
constrained through p modes \citep[e.g][and references
therein]{Howe2009}and so, in this paper, we assumed a rate similar
to the rest of the radiative zone. However, we note that g modes
would provide tighter constraints on the core rotation rate
\citep{Garcia2007, Mathur2008}.} such a large change cannot be ruled
out with current inversions of the solar rotation profile (see
Appendix \ref{section[appendix inversions]}). Even low-$l$ modes
carry so little information about the conditions of the core that
the uncertainties associated with any inversions are very large.

\begin{table}
\centering \caption{Variation in $\Omega/2\pi$ required to
explain a $0.15\,\rm\mu Hz$ change in the rotational splitting of
$l=1$ and 2, $n=17$ modes.}\label{table[splitting inversions]}
\begin{tabular}{cccc}
  \hline
  \noalign{\smallskip}
  Shell & Inner radius of  & $l=1$ & $l=2$\\
   & shell $(r/R_\odot)$ & $(\rm\mu Hz)$ & $(\rm\mu Hz)$ \\
  \noalign{\smallskip}
  \hline
  \noalign{\smallskip}
  core & 0.00 & 2.36 & 3.27\\
  radiative zone & 0.20 & 0.50 & 0.48\\
  convection zone & 0.70 & 0.47 & 0.47\\
  surface shear layer& 0.95 & 0.49 & 0.49\\
  \noalign{\smallskip}
  \hline
\end{tabular}
\end{table}

\subsection{The rotation profile of a solar-type star}

Asteroseismic data often consist of just a few months of
observations in which only the low-$l$ modes are visible. Thus
asteroseismic data are often similar to the 182.5\,d Sun-as-a-star
subsets analyzed here. We note here that this analysis is relevant
for main sequence stars for which no mixed modes or g modes can be
detected and used to aid inferences of the internal rotation rate.
Therefore it is difficult to make inversions of the internal
rotation profile of a star. However, lightcurves of a star, such as
those observed by CoRoT and Kepler, can show rotational modulation
due to the presence of magnetic stellar activity, such as spots and
active regions. The modulation of the light curve can, therefore, be
used to determine the surface rotation rate of a star
\citep[e.g.][]{Mosser2005, Mosser2009, Mathur2010, Ballot2011},
which can then be compared to the mean internal rotation rate
determined from the oscillations. The sidereal rotation rate of the
Sun, as estimated from the rotation of
 sunspots at the surface, is approximately 27 d. The mean rotation rate of
 the solar interior, determined from low-l rotational frequency splittings,
 is 28\,d. However, if we were to
use the extreme values of the splittings obtained from a single
182.5\,d subset, similar to those available in asteroseismic
studies, we would find the mean rotation rate to be either
$22\pm3\rm\,d$ or $44\pm12\rm\,d$. One may, therefore, erroneously
conclude that the average rotation rate of the stellar interior is
different to the surface rotation rate.

\section{Discussion}\label{section[discussion]}

As with previous studies, \citep[e.g.][]{Jimenez2002,
Gelly2002, Garcia2004, Garcia2008, Salabert2011}, we find no 11-yr
solar cycle variation in the rotational splittings. Discernible
quasi-periodic mid-term $(\sim2\,\rm yr)$ variations are present in
the splittings. However, we show that these variations are due to
realization noise.

Many authors have discussed the biases associated with estimating
splittings from Sun-as-a-star data \citep[e.g.][]{Eff-Darwich1998,
Appourchaux2000, Chaplin2006, Garcia2008}. We have shown that,
despite taking the advice of \citet{Chaplin2006}, the realization
noise has a larger effect on rotational splittings than accounted
for by formal uncertainties, implying that the uncertainties on the
fitted splittings are underestimated. The effect of the realization
noise produces an erroneous apparent mid-term $(\sim 2\,\rm yr)$
signal in the observed splittings that could, potentially, look like
the quasi-biennial solar-cycle related signal that is observed in
the mode frequencies \citep{Broomhall2009, Salabert2009,
Fletcher2010}. Although we have shown that the mid-term signal is an
artifact associated with the realization noise and the data analysis
process the periodicities in the splittings may be mistakenly
interpreted as a physical variation in the rotation rate. Such
variations are unlikely because they would require either very large
changes in the core rotation or changes in the near surface that,
although smaller, are ruled out through observations of higher-$l$
modes.

It is not surprising that the splittings observed by BiSON and GOLF
are highly correlated, since they both are observing the same star.
In fact we expect the splittings to be correlated whether these
variations are of solar origin or whether, as we have shown, they
are due to realization noise. \citet{Jimenez2004a} found that there
was a strong degree of correlation between numerous mode parameters,
such as mode linewidths and power densities, observed in GOLF and
BiSON data. They concluded that this showed that the estimated mode
parameters were dominated by the same mode realization noise (the
signature of stochastic excitation). \citet{Howe2006} observed that,
after the solar-activity dependence had been removed from the
frequencies, significant fluctuations in the frequencies were highly
correlated between BiSON, GONG, and MDI data. \citeauthor{Howe2006}
explained these correlations in terms of the stochastic nature of
the mode excitation, which could be interpreted as a form of
realization noise. The high level of correlation observed here
between the GOLF and BiSON splittings is also due to the mode
realization noise. This supports the conclusion that the mode
realization noise is responsible for the observed variation in the
splittings, which again demonstrates how careful one needs to be,
particularly when data have correlated noise sources.

Although the spurious variations do not have a significant effect on
inferred rotation profiles of the Sun they could be important for
any asteroseismic inferences of the average internal rotation rate
of main sequence stars for which g modes and mixed modes have not
been detected. Firstly because, unlike for the Sun, we must rely on
the splittings estimated from low-$l$ modes only. Secondly, because
we often have only one relatively short data set, of the order of a
few months, from which to extract the splittings. An additional
consideration is that the accuracy of the determined splitting is
dependent on the size of the splittings relative to the width of the
modes i.e. whether the modes overlap in frequency \citep{Ballot2006,
Ballot2008}. We have shown that care must be taken in using average
splittings (extracted from data sets of a few months duration) to
infer the mean internal rotation rate (which, for example, might be
compared in asteroseismic studies to the surface rotation rate
inferred directly from rotational modulation of the light curve,
e.g. \citealt{Ballot2011}). Artificial data, such as those available
through AsteroFLAG \citep{Chaplin2008}, should be used to test for
potential biases in the results. Care should be taken when
interpreting the observed splittings of low-$l$ modes in both
astero- and helioseismic data.

\appendix\section{Inversion process: Forward calculations}\label{section[appendix
inversions forward]}

A rotation profile of the Sun was simulated, based on the average of
2D inversion profiles computed over 15\,yrs using a Regularized
Least Squares \citep[RLS, ][]{Schou1994} inversion method and Global
Oscillations Network Group data \citep[GONG,][]{Harvey1996,
Thompson1996, Howe2005}. The GONG instrument make spatially resolved
observations of the Sun and is thus able to observe much higher-$l$
modes than those examined here. The GONG observations were used to
build a latitudinally dependent rotation profile of the Sun between
0.5 and 1.0 solar radii $(R_\odot)$. Below $0.5\,R_\odot$ the
rotation was taken to be constant and equal to the mean inferred
rotation rate across all latitudes at $0.5\, R_\odot$. We varied the
rotation rate in four different regions of the solar interior; the
solar core $(0.00R_{\odot}\le r<0.20R_{\odot})$, the radiative zone
$(0.20R_{\odot}\le r<0.70R_{\odot})$, the convection zone
$(0.70R_{\odot}\le r<0.95R_{\odot})$, and the near-surface shear
layer $(0.95R_{\odot}\le r<1.00R_{\odot})$. The splittings were
calculated when the rotation rate in each region was varied between
$330\le\Omega/2\pi\le530\,\rm nHz$. The $l=1$ and 2 splittings,
$\delta\nu$, were then determined \citep[see for
example][]{Chaplin1999}.

\begin{figure}
\centering
  \includegraphics[width=0.4\textwidth, clip]{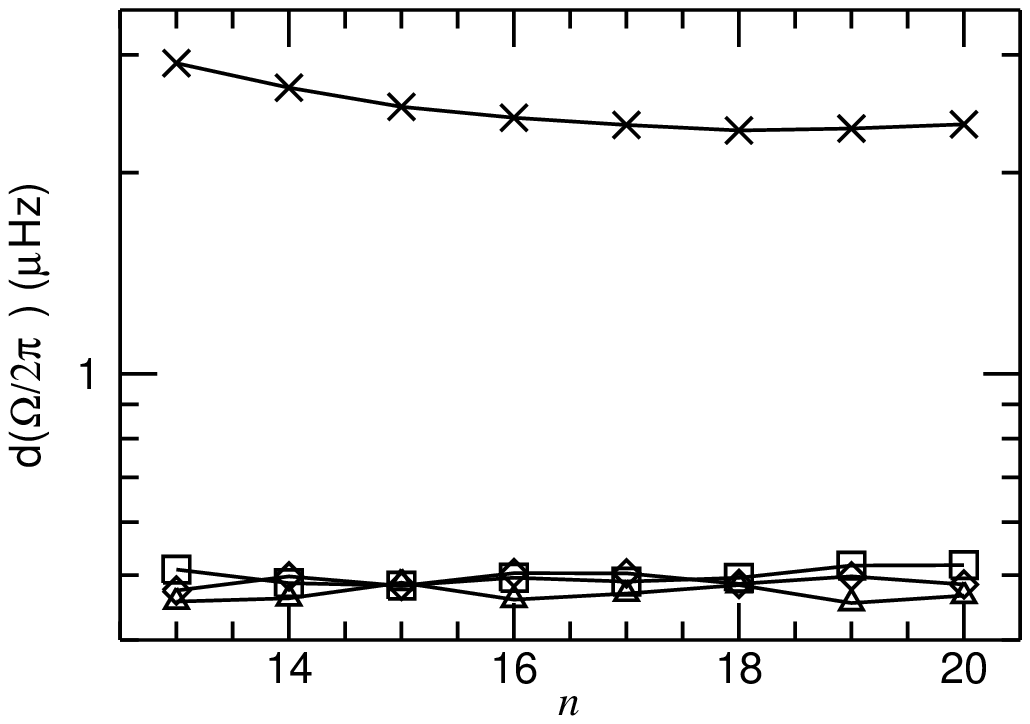}\\
  \includegraphics[width=0.4\textwidth, clip]{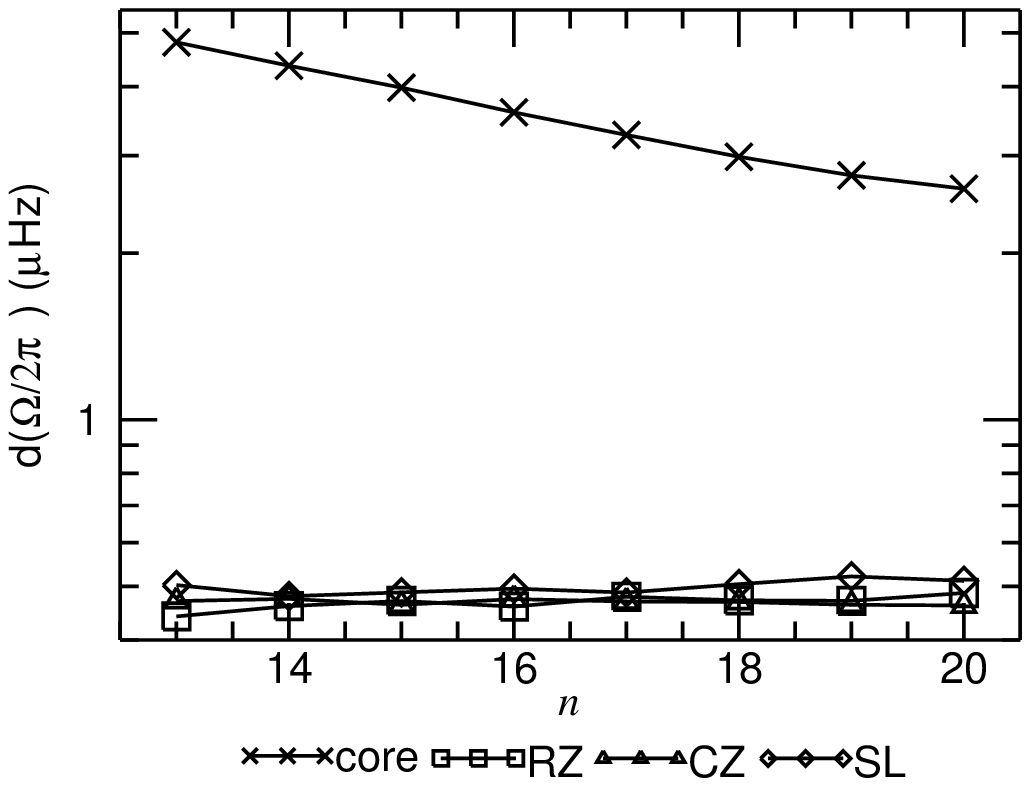}\\
  \caption{Change in rotation rate required to account for a
  $0.15\,\rm\mu Hz$ change in the observed splitting as a function of $n$,
  for $l=1$ (top panel) and $l=2$ (bottom panel) modes. The different symbols
  correspond to the results obtained when the rotation rate in a
  specific region was altered (see legend).}
  \label{figure[shell]}
\end{figure}

To first order, the magnitudes of the splittings vary linearly with
the rotation rate in a given region. That linear factor depends on
the position of the specific region in the solar interior and so can
be used to infer the change in rotation rate in a specific region
that is required to produce a change in splitting of $0.15\,\rm\mu
Hz$. These values are plotted in Fig. \ref{figure[shell]} and are
listed in Table \ref{table[splitting inversions]} for $l=1$, $n=17$
and $l=2$, $n=17$ (corresponding to modes at 2561 and $2621\,\rm\mu
Hz$ respectively). The results in Fig. \ref{figure[shell]} and Table
\ref{table[splitting inversions]} indicate by how much the rotation
rate in a particular region would have to deviate from the values
obtained from the 15-yr 2D RLS inversions to explain the observed
results. The required deviation in rotation rate change is similar
for the radiative zone, the convection zone and the surface shear
layer and is approximately constant across the range of $n$ examined
here. However, the change in rotation rate required in the core is
substantially larger and depends on $n$.

As expected from the rotation kernels, the observed splittings are
much more sensitive to surface regions than to the core. The changes
in Table \ref{table[splitting inversions]} can be compared to the
mean rotation rate at the surface, which is approximately
$\Omega/2\pi=430\,\rm nHz$. The rotation rate in the core would have
to be approximately 5 times faster than the mean solar surface
rotation rate to account for an increase in the splitting of
$0.15\,\rm\mu Hz$, while the rotation rate in the other regions of
the solar interior would have to change by more than 100\,per cent
of the mean solar value. In other words, even if the shell
responsible for the change in splitting is close to the surface, a
large, but not inconceivable, change in rotation rate is required to
explain a difference in the observed splitting of $0.15\,\rm\mu Hz$.

\section{Variations in the inferred solar rotation profile}\label{section[appendix inversions]}

\begin{figure}
\centering
  \includegraphics[width=0.4\textwidth, clip]{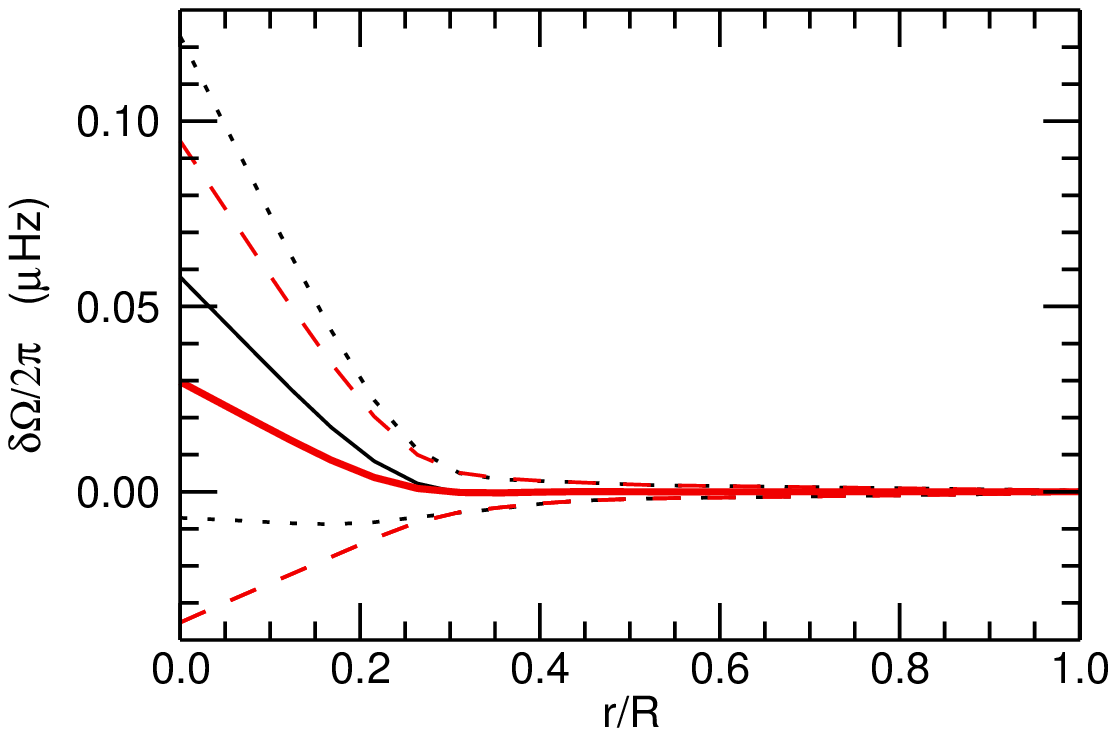}\\
  \includegraphics[width=0.4\textwidth, clip]{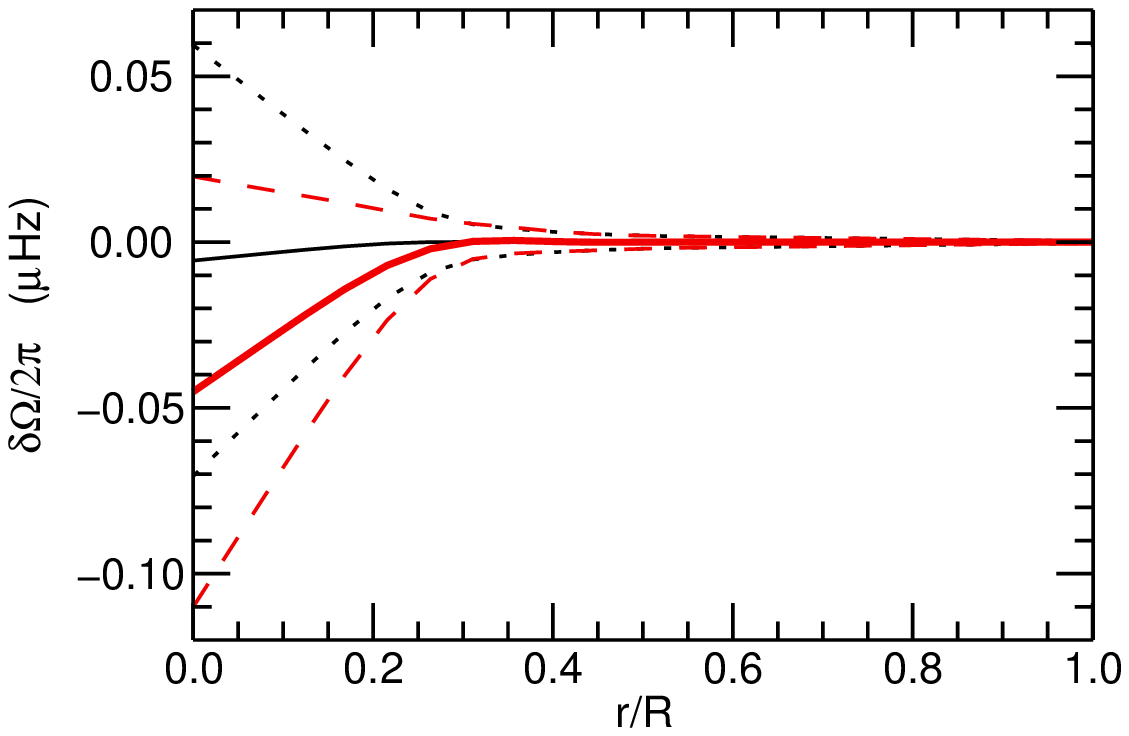}\\
  \caption{Residuals in the inferred profile at the equator
  once a mean profile was subtracted. The top panel shows the results
  obtained using the splittings observed in a 182.5\,d subset that began on 2001 April 10
  while the bottom panel shows results using the splittings observed in the 182.5\,d subset
   that began on 2008 October 7. The black solid line shows the results obtained
   using the BiSON data, while the $1\sigma$ uncertainties are denoted by the black dotted lines.
   The red thick line shows the results obtained using the GOLF data,
   while the $1\sigma$ uncertainties are denoted by the red dashed lines.}
  \label{figure[inversions selected]}
\end{figure}

We have combined the low-degree splittings observed in the BiSON and
GOLF data with a set of intermediate-degree splittings made from the
average of three 72\,d Michelson Doppler Imager
\citep[MDI,][]{Scherrer1995} sets observed between 1996 and 1997.
Each set of splittings was then inverted using 2D RLS methods
\citep{Schou1994}. Fig. \ref{figure[inversions selected]} shows
residuals in the inferred rotation profile at the equator once a
mean profile has been subtracted. In regions where no information is
available (i.e. the core) the 2D RLS inversion method extrapolates a
solution. In these regions the averaging kernels, the functions that
describe how the estimate of the solution corresponds to a spatial
average of the underlying true solution, are not well localized. The
rotation profiles plotted in Fig. \ref{figure[inversions selected]}
were generated using two sets of splittings measured by BiSON and
GOLF. These were chosen because they correspond to epochs when the
observed splitting deviates noticeably from the mean. In each case
the rotation profiles differ from the mean, and from each other, in
the deep interior but these differences are less than the formal
uncertainties for the inversions. Therefore, the spurious changes in
the observed splittings do not have a significant effect on the
inferred rotation profile of the Sun. This is not entirely
unexpected because even though the (model-dependent) spatial
weighting functions (kernels) of low-$l$ modes do probe the solar
core, they are still more sensitive to the surface regions. Indeed,
the sound speed in the deep interior is much larger than near the
 surface, so the time the modes dwell in the core is small relative to the
 outer regions.

\section*{Acknowledgements}

We thank the referee for their insightful comments. This paper utilizes data collected by the Birmingham
Solar-Oscillations Network (BiSON), which is funded by the UK
Science Technology and Facilities Council (STFC). We thank the
members of the BiSON team, colleagues at our host institutes, and
all others, past and present, who have been associated with BiSON.
The GOLF instrument on board \emph{SOHO} is a cooperative effort of
many individuals, to whom we are indebted. \emph{SOHO} is a project
of international collaboration between ESA and NASA. A.M.B., W.J.C., and
Y.E. acknowledge the financial support of STFC. D.S.
acknowledges the support from the Spanish National Research Plan
(grant PNAyA2007-62650) and from CNES. R.A.G. thanks the support of
the CNES/GOLF grant at the CEA/Saclay. R.H. thanks the National Solar Observatory for computing resources. We thank R. New for helpful comments and discussions. NCAR is supported by the National Science Foundation.

\bibliographystyle{mn2e_new}
\bibliography{splittings_paper_v1}

\end{document}